\documentstyle[pra,aps]{revtex}

\newcommand{\be}{\begin{equation}}
\newcommand{\ee}{\end{equation}}
\newcommand{\ba}{\begin{eqnarray}}
\newcommand{\ea}{\end{eqnarray}}
\newcommand{\lab}{\label}
\newcommand{\re}{\ref}

\begin{document}

\twocolumn[
\title{Relationships among coefficients in deterministic and stochastic 
transient diffusion}
\author{Z. Kaufmann}
\address{Institute for Solid State Physics,
E\"otv\"os University,
M\'uzeum krt. 6-8, H-1088 Budapest, Hungary}
\maketitle

\mediumtext
\begin{abstract}%
Systems are studied in which transport is possible due to large extension
with open boundaries in certain directions but the
particles responsible for transport can disappear from it by 
leaving it in other directions, by chemical reaction or by adsorption.
The connection of the total escape rate, the rate of the disappearance
and the diffusion constant is investigated.
It leads to the observation that the diffusion coefficient defined by
$\langle x^2\rangle$ is in general different from the one present in the
effective Fokker-Planck equation.
The result makes it possible to generalize
the Gaspard-Nicolis formula for deterministic systems to this
transient case.
\end{abstract}
] 

\narrowtext

Randomness and diffusion are common features of extended stochastic and
chaotic systems\cite{Ha,LiLi,BuSi,MaZw,GeNi,ScFrKa,GrFu,KlDo,Ga}.
Among deterministic systems diffusion has been much studied in the Lorentz
gas \cite{BuSi,MaZw}.
As more simple models proper 1D maps \cite{GeNi,ScFrKa,GrFu,KlDo} and a 2D
map \cite{Ga} has been introduced, which were built as chains of maps.
Ref.\ \cite{GaNi} has shown a relationship between the diffusion coefficient
and microscopic quantities, namely the Liapunov exponent and the
Kolmogorov-Sinai entropy referring to the repeller.
This was later generalized to other transport coefficients \cite{DoGa} and
case of small external field \cite{TVoBr}.
The Liapunov exponent was independently calculated for the random Lorentz
gas \cite{BeDo}.

There are systems in which particles can escape in directions transversal
to the extension of the system
raising interestic problems \cite{KLNSz} in the field of transient chaos
\cite{T96} in particular in the critical case \cite{NSz,LSz}.
A simple example is a channel in a mesoscopic system
modeled by a strip of Lorentz gas with open side boundaries.
Another example is
the Troll-Smilansky model for chaotic scattering consisting of a
1D infinite periodic array of soft potential valleys,
which is a system related to a model of ionization \cite{TrSm,LaGr}.
Further example is the motion in an infinite set of resonances in the
phase space of Hamiltonian systems
when one considers the particle to escape when it leaves the set of chosen
resonances \cite{Da}.
Diffusion has also been investigated on a chaotic saddle
in a model for the interaction of a particle with electrostatic
wavepacket \cite{DaKa}.
Particles can also be lost from the point of view of diffusion by
absorption, chemical reaction \cite{Pr} or by other ways.
Particles in such transiently chaotic or stochastic systems can
diffuse in the extended direction or directions for some
time and then escape either through the ends in the extended direction
(if the system is finite) or in the other directions or other ways.
Therefore the average period for which they take part in transport is
finite, and remains finite even in the limit when the size of the system in
the extended direction goes to infinity.

The aim of the present paper is to generalize the Gaspard-Nicolis formula
\cite{GaNi} to the case of the above described transient diffusion
in deterministic systems.
For this purpose it is studied how the total escape rate separates to
terms related to the extended and the transversal direction.
This investigation is made in a general way leading to interesting results
for both deterministic and stochastic systems.
For simplicity the system is assumed to possess one extended direction
with a discrete translational symmetry, and in
most of the considerations here, an inversion symmetry that reverses the
extended direction.
For sake of convenience the primitive cells with respect to the
translational symmetry
shall be labeled by a discrete variable $x$ that is monotonous in the
extended direction.
The rest of variables specifying the state of the particle shall
be assembled in $y$.
It is convenient to show the choice of $x$ and $y$ in case of the strip of
Lorentz gas.
Here, as in general, it is easier to study the system in discrete instead
of continuous time.
Taking the surfaces of the disks as Poincar\'e surface the state of the
particle on it can be given by the coordinates $x,q,\alpha$ and $\beta$.
By $x$ we mean the ordered label of the periods of the structure. $q$ is a
label of the disk inside one period, $\alpha$ is the angle of position on
the disk and $\beta$ is the angle of reflection.
Then $y$ corresponds to $(q,\alpha,\beta)$.

The general evolution equation for the probability density
$\phi_t(x,y)$ of the particle can be written as
\be
\phi_{t+1}(x,y)=\sum_{j=-J}^J \sum_{y'} w_{j,y,y'}\phi_t(x-j,y')\;, \lab{ev}
\ee
where, if $y$ contains continuous variables one can consider the sums
as integrals over the continuous components or one can use
coarse graining with arbitrary precision.
Here $\phi_t=0$ if the argument falls outside the region of the system.
The maximal jump $J$ can be assumed to be finite, or the transition
probability $w_{j,y,y'}$ to decay fast in $j$.
The translational symmetry is implied in the form of (\re{ev}), while an
inversion symmetry can be written as $w_{j,y,y'}=w_{-j,Ty,Ty'}$, where
$T^2 y=y$ for every $y$.

Two representative classes of such systems can one keep in mind here.
The first is a 2D random walk in a strip, for simplicity assuming no
memory, for which
\be
\phi_{t+1}(x,y)=\sum_{j=-J}^J \sum_{k=-K}^K W_{jk} \phi_t(x-j,y-k)
\lab{2d}
\ee
applies.
This can be considered as a rough description of the Lorentz gas strip.
The inversion symmetry can be a point inversion or a line inversion
symmetry.
To give an account of correlation between transitions 1D walks with memory
are taken as a second group of examples.
Then (\re{ev}) can be used conceiving $y$ as the memory containing, say,
$n$ number of past steps $j_1, j_2, \ldots j_n$ and
$w_{j_{n+1},y,y'}=P(j_{n+1}|j_n,j_{n-1},\ldots j_1)$ is the conditional
probability of the next step.
The inversion symmetry implies
$P(j_{n+1}|j_n,j_{n-1},\ldots j_1)=P(-j_{n+1}|-j_n,-j_{n-1},\ldots -j_1)$.

For general considerations we shall return to (\re{ev}) taken without
restriction to 1D.
For the diffusion process the long time behavior of the system is
important.
That is governed by the leading eigenfunction of the right hand side of
(\re{ev}), i.\ e.\ the solution $\phi_t$ for which
$\phi_{t+1}=e^{-\kappa}\phi_t$ and the total escape rate $\kappa$ is minimal.
This $\phi_t$ shall be called the asymptotic state.
The boundary of the system should also show the symmetry, so the region of
the system should be defined by independent
conditions in $x$ and $y$ ($x\in R_x$ and $y\in R_y$).
If the particle from a point $(x_0,y_0)$ jumps to a point $(x,y)$, for
which $y\not\in R_y$ the particle shall be considered to escape in $y$
direction, while in case $y\in R_y$  but $x\not\in R_x$ it escapes in $x$
direction.

To find the relation of the escape rates and the diffusion coefficient
is simple in certain cases but is problematic in general.
It is enlightening to study the simple cases first.
These are the ones in which the asymptotic solution
separates as a product $\phi(x,y)=\psi(x)\omega(y)$.
This happens if
the transition probability matrix is a sum of diadic products
$w_{jyy'}=\sum_s u_j^{(s)}v_{yy'}^{(s)}$ and the partial diffusion equations
in $x$ direction have a common leading eigenfunction, i.~e.\
$\sum_j u_j^{(s)}\psi(x-j)=\mu_s\psi(x)$ such that for every $s$
$\mu_s$ is the maximal eigenvalue.
This property shall be denoted by $S_x$.
It is easy to see that in this case the eigenvalue equation in $y$
direction
$\sum_{y's}v_{yy'}^{(s)}\mu_s\omega(y')=e^{-\kappa}\omega(y)$
determines $\omega$ and $e^{-\kappa}$.
An example for this case among the 2D walks (\re{2d}) is a walk on a
square lattice $[1,L]\otimes[1,M]$ with
\be
\{W_{jk}\}_{-1,-1}^{1,1}\!\!=\!\!
\left(\begin{array}{ccc}0&q&0\\0&r&0\\p&0&p\end{array}\right)\!\!=\!\!
\left(\!\!\begin{array}{c}q\\r\\0\end{array}\!\!\right)\!\!\circ\!
(010)\!+\!\!
\left(\!\!\begin{array}{c}0\\0\\p\end{array}\!\!\right)\!\!\circ\!
(101)
\ee
whose solution reads\newline
$\phi=\sin\left(\pi\frac{x}{L+1}\right)
\left[2\frac{p}{q}cos\left(\frac{\pi}{L+1}\right)\right]^{y/2}
\sin\left(\pi\frac{y}{M+1}\right)$.

Another possibility is when the eigenmode is common in $y$ direction,
i.\ e.\ $\sum_{y'} v_{yy'}^{(s)}\omega(y')=\nu_s\omega(y)$
(property $S_y$).
Then $\sum_{js}u_j^{(s)}\nu_s\psi(x-j)=e^{-\kappa}\psi(x)$ determines
$\psi$ and $e^{-\kappa}$.
When the properties $S_x$ and $S_y$ are both
satisfied then $e^{-\kappa}=\sum_s \mu_s\nu_s$.

To obtain an evolution equation of the distribution on large scale in $x$
direction in either of the above cases with $\phi(x,y)=\psi(x)\omega(y)$
one has to study the evolution of $\psi$. It is convenient to choose the
normalization $\sum_y\omega(y)=1$, since then $\psi(x)=\sum_y\phi(x,y)$.
Using this equation and (\re{ev}) one obtains
\be
\psi_{t+1}(x)=\sum_j\tilde w_j\psi_t(x-j)\;,\;\;
\tilde w_j\equiv\sum_{yy'}w_{j,y,y'}\omega(y')\;.
\ee
Obviously the probability that a particle at $x_0=x-j$ does not escape in $y$
direction in the next step is
$e^{-\kappa_y}=\sum_j\tilde w_j\equiv\sum_{jyy'}w_{jyy'}\omega(y')$,
which is independent of $x_0$.
One can separate this escape with substitution
$\psi_t(x)=e^{-\kappa_y t}g_t(x)$, which yields
$g_{t+1}(x)=\sum_j\tilde w_j e^{\kappa_y}g_t(x-j)$.
Since this equation describes a random walk and the system is extended in
$x$ direction an effective Fokker-Planck equation
$g_{t+1}(x)=g_t(x)+D_{FP}g_t''(x)$
is valid on large scales.
Returning to $\psi$ it takes the form
\be
\psi_{t+1}(x)=\left(1+D_{FP}\frac{d^2}{dx^2}\right)e^{-\kappa_y}\psi_t(x)\;.
\lab{fp}
\ee
Its dominant solution in case of a channel of length $L$ is
\be
\psi_t(x)=e^{-\kappa t}\sin\left(\pi\frac{x}{L}\right)\;,\;\;
\kappa=\kappa_x+\kappa_y\;,\lab{sols}
\ee
and $\kappa_x=(\pi^2/L^2)D_{FP}+{\cal O}(L^{-3})$,
if $\kappa_x$ is defined such, that
$e^{-\kappa_x}$ is the conditional probability that a point does not escape in
$x$ direction if it does not escape in $y$ direction.
Note, that
$\kappa_y$ may also depend on $L$.

The general case, when $\phi(x,y)\neq\psi(x)\omega(y)$ is more
complicated.
The main point is, that $\kappa_y$ becomes dependent on $x$ and $L$,
the length of the system in $x$ direction.
However, in case when $L$
is much larger than the size in other directions
and there is an inversion symmetry
it shall be shown, that
apart from a vicinity of the ends of the channel
the deviation of $\kappa_y$ from a value $\kappa_y^{(\infty)}$ 
is proportional to $f''(x)/f(x)$. 
Here $f(x)$ is introduced analogously to $\psi(x)$ as
$f(x)=\sum_y\phi(x,y)$ and 
$\kappa_y^{(\infty)}$ is the value of $\kappa_y$ for the homogeneous
solution in case $L=\infty$.
This makes it possible to write down a proper effective Fokker-Planck
equation.
It is suitable to 
choose a segment of the system separated by $x=\rm const$ planes,
such that its size is still much larger than the transversal size, but
much smaller than $L$.
Such a segment feels values $f(x_1)$, $f(x_2)$ of $f$ at its ends with a
common exponential decay $e^{-\kappa t}$. It is assumed that the diffusion
in the system mixes the contribution of sites with different $y$
coordinates.
So systems are excluded in which the particles
from one site can not fill the whole system,
thereby different initial distributions can lead to different asymptotic
states.
With this assumption the effect of the transversal distribution
$\phi(x,y)/f(x)$ decays fast from the ends of the segment towards its inside.
On the other side, in the asymptotic state the transversal distribution at
the ends of the segment is almost identical to the one in the middle.
Therefore one can expect the distribution in the inside of the segment is
determined by the values $f(x_1)$, $f(x_2)$ and the decay rate $\kappa$.

Thereby the distribution can have three free parameters.
Since the evolution equation (\re{ev}) is linear in the distribution, the
asymptotic distributions $\phi$ and $f$ can be assumed to have linear
combination form
\ba
\phi(x,y)&=&[\phi^{(\infty)}(x-x_0,y)+c_1\phi^{(1)}(x-x_0,y)\nonumber\\
&&+c_2\phi^{(2)}(x-x_0,y)]e^{-\kappa t}\;,\\
f(x)&=&[1+c_1 f^{(1)}(x-x_0)+c_2 f^{(2)}(x-x_0)]e^{-\kappa t}\;,
\ea
where $f^{(k)}(x)=\sum_y\phi^{(k)}(x,y)$ for $k=1,2$ and $x_0$ is the
center of the segment.
Here $\phi^{(\infty)}(x,y)=\omega^{(\infty)}(y)$ is the solution in the
limit $L\rightarrow\infty$ that is independent of $x$.
The corresponding escape rate shall be denoted by $\kappa_y^{(\infty)}$.
$\phi^{(1)}$ is the asymptotic solution of (\re{ev}) with antisymmetric
boundary conditions $f(x_1)=-f(x_2)$.
This term is responsible for current through the middle of the segment.
Starting with a $\phi^{(\infty)}$ alone and symmetric boundary conditions
$f(x_1)=f(x_2)=a e^{-\kappa t}$ one observes in general that $\phi$ in the
middle decays faster or slower than on the boundary depending on the sign
of $\kappa-\kappa^{(\infty)}$.
In the asymptotic state this leads to a hump- or vale-shape term $f^{(2)}$
in $f$, which corresponds to some $\phi^{(2)}$.
The partial distributions can be approximated as $f^{(1)}(x-x_0)=x-x_0$
and $f^{(2)}(x-x_0)=(x-x_0)^2-b$ in a vicinity of the middle of the
segment if $\phi^{(k)},\;k=1,2$ are properly
normalized.
This means, that the general solution is
\ba
\lefteqn{\phi(x,y)=f(x_0)\phi^{(\infty)}(x-x_0,y)}\nonumber\\
&&\;\;\mbox{}+f'(x_0)\phi^{(1)}(x-x_0,y)\nonumber\\
&&\;\;\mbox{}+\frac{f''(x_0)}{2}(\phi^{(2)}(x-x_0,y)+b\phi^{(\infty)}(x-x_0,y))
\lab{adi}
\ea
in a vicinity of the point $x_0$.
The local rate of escape $\kappa_y(x_0)$ in $y$ direction in the middle
can be calculated as
\ba
1-e^{-\kappa_y(x)}&=&\frac{E(x_0)}{f(x_0)}\;,\lab{ky}\\
E(x_0)&=&\sum_y\phi(x_0,y)-\sum_{yy'}w_{jyy'}\phi(x_0,y')\;,\lab{ex}
\ea
where $E(x_0)$ is the flow of escape in $y$ direction at $x_0$.
Then clearly
\be
E(x_0)=E_{\infty} f(x_0) \!+\!E_1 f'(x_0)
\!+\!\frac{(E_2\!+\!bE_{\infty})}{2}f''(x_0),\lab{e+}
\ee
where $E_k, k=\infty,1,2$ are constants characteristic of $\phi^{(k)}$,
respectively.
It can be shown that $E_1=0$. To see this one can separate Eq.\ (\re{ev})
to terms related to transition from inside of the chosen segment and from
outside.
Conceiving $\phi_t(x,y)$ as a vector $\Phi=\{\Phi_i\}_i$, where any value
of the index $i$ corresponds to a point in $(x,y)$ space inside the
segment and assuming the asymptotic state (i.~e.\
$\phi_{t+1}=e^{-\kappa}\phi_t$) one obtains
$e^{-\kappa}\Phi_i=\sum_j\tilde w_{ij}\Phi_j+\chi_i$.
Here $\tilde w$ describes the transitions inside, $\chi$ the
transitions from outside, and $\tilde w, \chi$ can be constructed using
$w$ and values of $\phi$ outside but near the boundary of the segment.
The solution for $\Phi$ is
$
\Phi=\left(e^{-\kappa}I-\tilde w\right)^{-1}\chi\;,
$
where $I$ is the unit matrix.
The boundary conditions are antisymmetric for the symmetry transformation $T$ in 
the state $\phi^{(1)}$, thereby $\chi$ is also antisymmetric.
$w$ and $\tilde w$ are symmetric for $T$.
Consequently $\Phi^{(1)}$ and the corresponding $\phi^{(1)}$ are
antisymmetric.
Therefore (\re{ex}) yields $E_1$=0.
Using (\re{ky}) and (\re{e+}) one obtains
\be
\kappa_y(x)=\kappa_y^{(\infty)}+\eta\frac{f''(x)}{f(x)}+{\cal O}(L^{-3})
\lab{kyx}
\ee
with a suitable constant $\eta$, since for large $L$ one expects
$f''={\cal O}(L^{-2})$.
This is not valid near the ends of the channel where the middle of the
segment can not be placed.
Then the analog of the effective Fokker-Planck equation (\re{fp}) becomes
of the from
\be
f_{t+1}(x)=\left(1+D_{FP}\frac{d^2}{dx^2}\right)e^{-\kappa_y(x)}f_t(x)\;,
\lab{fp2}
\ee
or, with substitution of (\re{kyx})
\be
f_{t+1}(x)=e^{-\kappa_y^{(\infty)}}
\left(1+(D_{FP}-\eta)\frac{d^2}{dx^2}\right)f_t(x)\;.\lab{fp3}
\ee
It is plausible to assume that a distribution corresponding to (\re{adi})
sets in earlier in time than the asymptotic state
in the extended direction.
Then (\re{fp2},\re{fp3}) are also valid for general $f(x)$ distributions
still not in the asymptotic state.
(\re{fp3}) shall be applied later for such a case, but here the asymptotic
solution is important.
It is given by
\be
f_t(x)=e^{-\kappa t}\cos\left
[\sigma\left(\frac{x}{L}-\frac{1}{2}\right)\right]\;,\lab{sol}
\ee
with
$\kappa=\kappa_y^{(\infty)}-
\log[1-(D_{FP}-\eta){\sigma^2}/{L^2}]$,
which is valid inside the channel with a deviation at its ends, such that
$f_t(x)$ reaches zero at the ends ($x=0$ and $x=L$), while (\re{sol})
takes zero value at $\xi=(1-\pi/\sigma)L/2$.
In the limit $L\rightarrow\infty$ the value of $\xi$ becomes constant, since
the neighborhood of the channel behaves in the same way in this limit.
Therefore $\sigma=\pi+{\cal O}(L^{-1})$.
Using (\re{kyx}) and (\re{sol}) one can notice, that $\kappa_y(x)$ is
constant apart from the vicinity of the ends of the channel.
Its value is
\be
\kappa_y^{(L)}=\kappa_y^{(\infty)} -\frac{\pi^2}{L^2}\eta+{\cal
O}(L^{-3})\;,\lab{kyl}
\ee
where the notation emphasizes its dependence on $L$.
Using this equation and the ones for $\kappa$ and $\sigma$
one obtains
the relation of the escape rates with the diffusion coefficient
\be
\kappa=\frac{\pi^2}{L^2}D_{FP}+\kappa_y^{(L)}+{\cal O}(L^{-3})\;.\lab{kl}
\ee
Note, that $(\pi^2/L^2)D_{FP}$ does not give $\kappa_x$ in general,
contrary to (\re{sols}).
Following \cite{GaNi} in case of deterministic process the
total escape rate can be related to the Liapunov exponents and the
Kolmogorov-Sinai entropy of the repeller, namely,
$\kappa=\sum_{\lambda_i>0}\lambda_i-h_{KS}$ \cite{EcRu,KaGr}.
This yields the generalization of the Gaspard-Nicolis formula \cite{GaNi}
\be
\sum_{\lambda_i>0}\lambda_i-h_{KS}=
\frac{\pi^2}{L^2}D_{FP}+\kappa_y^{(L)}+{\cal O}(L^{-3})\;.\lab{gng}
\ee

One more important question is whether $D_{FP}$ defined as the coefficient in
(\re{fp2}) is equal to the one defined by the mean square deviation of
$x$ as
$\int f_t(x)(x-x_0)^2\,dx/\int f_t(x)\,dx\propto 2D_{msd}t$,
starting from a state concentrated in  a vicinity of $x_0$.
Introducing $g_t(x)=e^{\kappa_y^{(\infty)}t}f_t(x)$
in Eq.\ (\re{fp3}) one can eliminate the factor containing
$\kappa_y^{(\infty)}$.
Then the equation becomes an evolution equation of a diffusion process
whose diffusion coefficient
\be
D_{msd}=D_{FP}-\eta\lab{dmsd}
\ee
is clearly equal to the
diffusion coefficient defined for $f_t(x)$ by mean square deviation.
So we can see $D_{FP}$ and $D_{msd}$ are in general different.
Comparing (\re{gng}) to the Gaspard-Nicolis formula \cite{GaNi}
one can notice that $\kappa_y^{(L)}$ has appeared as an additional term,
but using (\re{kyl}) and 
(\re{dmsd}) one can get another form of (\re{gng}) in which
$D_{msd}$ and $\kappa_y^{(\infty)}$ are present instead of
$D_{FP}$ and $\kappa_y^{(L)}$:
\be
\sum_{\lambda_i>0}\lambda_i-h_{KS}=
\frac{\pi^2}{L^2}D_{msd}+\kappa_y^{(\infty)}+{\cal O}(L^{-3})\;.
\ee

Numerical calculations have been made to test the validity of the
statements for $\kappa_y(x)$ and Eqs.\ (\re{fp2},\re{dmsd}) in two concrete
models.

A, The first model is a 2D random walk on a
square lattice $[1,L]\otimes[1,M]$ with
\be
\{W_{jk}\}_{-1,-1}^{1,1}=
\left(\begin{array}{ccc}0&0&p\\q&r&q\\p&0&0\end{array}\right)
\ee
and $p=0.1$, $q=0.2$, $r=0.4$.

B, The second model is a 1D random walk with two step memory such that
$P(j_{t+1}|j_t,j_{t-1})=R_{j_tj_{t-1}}Q_{j_{t+1}j_t}$
and $R_{+1}=0.45$, $R_{-1}=0.9$, $Q_{+1}=4/9$, $Q_{-1}=5/9$.
To ensure symmetry $Q$ and $R$ depend only on the product in their
subscript.
$Q_{+1}+Q_{-1}=1$, therefore $R$ describes the probability that the
particle does not escape in $y$ direction and $Q$ describes the relative
probability of the steps $j_{t+1}$.
A possibility was given to rest
for one step with a probability $g=0.01$ without changing $j_t,j_{t-1}$.

In model A $\kappa_y(x)$ was found to decay exponentially to
$\kappa_y^{(L)}$ coming from each of the ends towards the inside of the
region.
In model B $\kappa_y(x)$ is constant for $x=2,3,\ldots L-1$ and has
different values only at the endpoints.
In both models $\kappa$ and $\kappa_y(x)$ have been measured.
$D_{FP}$ has been calculated from them by (\re{kl}) for large $L$ and
$\eta$ from 
the $L$-dependence of $\kappa_y^{(L)}$ by (\re{kyl}).
$D_{msd}$ has been determined independently measuring
$\langle(x-x_0)^2\rangle$ for a well concentrated initial distribution.
The results satisfy (\re{dmsd}) up to the expected precision (5 digits).
In case of model A $D_{FP}$ has also been calculated by
$i(x)=D_{FP}(f(x)-f(x+1))e^{-\kappa_y(x)}$, where $i(x)$ is the current between
sites at $x$ and $x+1$. The results of this match the ones by (\re{kl}),
thereby confirming the validity of (\re{fp2}) and (\re{kl}), which was
used in (\re{gng}).

The author would like to thank P. Sz\'epfalusy
for valuable discussions
and Z. R\'acz for useful remarks.
This work has been supported in part by the 
German-Hungarian Scientific and Technological Cooperation 
{\em  Investigation of classical and quantum chaos},
by the Hungarian National Scientific Research Foundation under Grant 
Nos.\ OTKA T017493 and OTKA F17166,
the US-Hungarian Science and Technology Joint Fund in cooperation
with the NSF and the Hungarian Academy of Sciences under project
No.\ 286.

\end{document}